\documentclass{aa}  
\usepackage{graphicx}
\usepackage{txfonts}
\usepackage{natbib}
\newcommand{\bz}{\ensuremath{\langle B_z\rangle}}
\newcommand{\bs}{\ensuremath{\langle \vert B \vert \rangle}}
\newcommand{\nz}{\ensuremath{\langle N_z\rangle}}

\newcommand{\te}{\ensuremath{T_{\mathrm{eff}}}}
\newcommand{\mwd}{{\rm WD\,2047+372}}
\newcommand{\esp}{{\rm ESPaDOnS}}
\begin{document}

   \title{Discovery of an extremely weak magnetic field in the white dwarf LTT 16093 = \mwd\thanks{Based on observations made with the William Herschel Telescope operated on the island of La Palma by the Isaac Newton Group in the Spanish Observatorio del Roque de los Muchachos of the Instituto de Astrofisica de Canarias, and on observations obtained at the Canada-France-Hawaii Telescope (CFHT) which is operated by the National Research Council of Canada, the Institut National des Sciences de l'Univers of the Centre National de la Recherche Scientifique of France, and the University of Hawaii.}
}

   \author{J. D. Landstreet
          \inst{1,2}
          \and
          S. Bagnulo \inst{1}
          \and 
          A. Martin  \inst{1,3}
          \and
          G. Valyavin \inst{4}
          }

   \institute{Armagh Observatory, College Hill, Armagh BT61 9DG,
              United Kingdom
         \and
              Department of Physics \& Astronomy, University of Western
              Ontario, London, Ontario N6A 3K7 Canada\\
              \email{jlandstr@uwo.ca}
         \and
              Astrophysics Group, Keele University, 
              Staffordshire ST5 5BG, United Kingdom
         \and
              Special Astrophysical Observatory, Nizhnij Arkhiz, 
              Zelenchukskij Region, 369167 Karachai-Cherkessian Republic,
              Russia
             }

   \date{Received February 28, 2016; accepted May 1, 2016}

 
  \abstract
  {Magnetic fields have been detected in several hundred white dwarfs,
    with strengths ranging from a few kG to several hundred MG. Only a
    few of the known fields have a mean magnetic field modulus below
    about 1\,MG. }
  {We are searching for new examples of magnetic white dwarfs with
    very weak fields, and trying to model the few known examples. Our
    search is intended to be sensitive enough to detect fields at the
    few kG level. }
  {We have been surveying bright white dwarfs for very weak fields
  using spectropolarimeters at the Canada-France-Hawaii telescope, the
  William Herschel telescope, the European Southern Observatory, and
  the Russian Special Astrophysical Observatory. We discuss in
  some detail tests of the WHT spectropolarimeter ISIS using the known
  magnetic strong-field Ap star HD~215441 (Babcock's star) and the 
  long-period Ap star HD~201601 ($\gamma$\,Equ).  }
  {We report the discovery of a field with a mean field modulus of about
    57\,kG in the white dwarf LTT\,16093 = \mwd. The field is
    clearly detected through the Zeeman splitting of H$\alpha$ seen in
    two separate circularly polarised spectra from two different
    spectropolarimeters. Zeeman circular polarisation is also
    detected, but only barely above the $3 \sigma$ level. }
   {The discovery of this field is significant because it is the third
     weakest field ever unambiguously discovered in a white dwarf,
     while still being large enough that we
     should be able to model the field structure in some detail with 
     future observations.  }


   \titlerunning{Discovery of the very weak magnetic field of WD\,2047+372}

   \maketitle

%

\section{Introduction}

Magnetic fields in stars are increasingly recognised as a source of
significant physical effects (such as transfer of angular momentum and
influence on accretion and mass loss), and as a phenomenon affecting
other physical processes (such as convection and internal shear). On
the basis of many recent observations, we now know something about the
occurrence of magnetic fields in most of the main stages of stellar
evolution, from the pre-main sequence
\citep{DonaLand09,Huss12,Alecetal13} through the main sequence
\citep{DonaLand09, Petietal13, Wadeetal15}, the red giant
\citep{Aurietal15} and asymptotic giant \citep{Grunetal10} stages, and
finally the white dwarf stage \citep{Kepletal13,Ferretal15}.

In general, the magnetic fields of non-degenerate stars with low effective
temperatures ($\te\ \la 6500-7000$\,K) have fields driven by current
dynamo action, analogous to the situation in the Sun. Fields are
generally stronger in relatively rapid rotating stars than in more slowly
rotating stars, and the fields often have complex surface structure which
changes on a rather short timescale. 

Magnetic fields are found only in a few percent of stars hotter than
$\sim 7000$\,K, and they do not seem to be powered by stellar rotation.
In fact, the strongest fields are found on stars that are unusually
slowly rotating stars. The fields of hotter stars have a considerably
simpler structure than those of cooler stars, usually of qualitatively
dipolar topology, and their structure does not change significantly on
timescales of decades.  The magnetic fields of white dwarfs are
generally of this second (``fossil'') variety.

Although we now have observational information about what kinds of
fields occur in various stages of evolution, the physics underlying the
transformations of field strength and structure during long evolutionary
stages (such as the main sequence or the white dwarf state) or between
successive stages (such as from the red giant to the white dwarf
state) are far from clear. A particularly challenging aspect of this
general question is to understand the processes that occur as a
stellar field evolves from fossil to dynamo or vice versa. An
important goal of current magnetic studies is to try to shed light on
this evolution.  As a particular example, the fossil fields found in a
few percent of white dwarfs are thought to originate in an earlier
evolution stage of the star, and thus these fields potentially carry
useful information about the processes that occur as a magnetic white
dwarf forms from the preceding giant stage, or perhaps as it forms as
a result of a merger process in a close binary system sometime during
that giant stages \citep{ValyFabr99,Tout08,Wicketal14}.

Observationally, the kinds of information that may potentially
illuminate how magnetic fields developed and evolved after the red
giant stage are (1) the surface structures of the magnetic fields of a
suitable sample of white dwarfs, and (2) the frequency and
distribution of field strengths found in white dwarfs as functions of 
age, mass, and surface composition.

At present, several hundred magnetic white dwarfs have been
identified, mostly by observation of magnetic (Zeeman) splitting of
spectral lines. Many of the known magnetic white dwarfs have been
found recently from the Sloan Digital Sky Survey
\citep[SDSS,][]{Kepletal13}.  This survey (and most other surveys for
white dwarfs) have been carried out using low resolution (and
frequently low S/N) spectroscopy. With such data it is not possible to
recognise magnetic splitting in white dwarfs with fields less than
about 1\,MG.  However, it is known that white dwarf magnetic fields as
weak as a few kG occur \citep[e.g.,][]{Aznaetal04}.  The total range
of field strengths is thus more than four dex, from below 20\,kG to
about 800\,MG \citep{Ferretal15}.

In contrast to the high-field regime of the white dwarf field strength
distribution, very little is known about the weak-field limit. If we
measure the typical strength of the field by the mean field modulus
\bs, an average of the scalar field over the visible hemisphere of
the star, only about a dozen white dwarfs are known to have magnetic
fields with $\bs\ \la 300$\,kG. Consequently the field strength
distribution over mass, age, and composition in the weak-field limit
is very poorly constrained. The weakest fields now known are certainly
right at the detectability threshold. It is therefore not known
whether the detected weak-field white dwarfs are near the lower limit
of the complete field strength distribution, or whether this
distribution continues to still weaker fields that are currently
undetectable.  The single piece of data that we have on the regime
below 1~kG is the observation that $|\bz| \la 250$\,G ($3\sigma$
limit) in the bright DA white dwarf 40\,Eri\,B \citep{Landetal15}.

Apart from the uncertainty of their actual incidence,
the morphologies of the known weak-fields in WDs have never been
modelled in detail -- only a couple of stars have even been modelled
with the simplest structures that might be appropriate, for example a
magnetic dipole centred on the star but possibly inclined to the
rotation axis \citep{Valyetal05,Valyetal08,Landetal12}.

We have been carrying out a search for extremely weak fields among
bright ($V \leq 13.5$) white dwarfs, using spectropolarimetry and
high resolution spectroscopy, both of which are sensitive to kG
fields, far below the 1\,MG detection threshhold of low resolution
spectroscopy. In the course of this survey, we have discovered an
extremely weak magnetic field in the DA3.4 white dwarf LTT\,16093 =
\mwd. The field of this star has $\bs \approx 57$\,kG. Only two
magnetic white dwarfs are known that are definitely magnetic but have
fields for which \bs\ falls below this value.  Thus this star is an
important addition to a very small sample. In this paper we report the
details of this discovery, and describe what we can deduce about the
star from the data obtained so far.

\section{Observations}

As discussed in detail by \citet{Koesetal98,Koesetal09,Landetal15},
searches for weak fields in white dwarfs using unpolarised
spectroscopy can detect fields as low as $\bs \sim 10 - 20$\,kG
provided the spectrograph has resolving power of $R \ga 20\,000$.
Still weaker fields can be detected with spectropolarimetry, measuring
\bz, the line-of-sight component of the magnetic field averaged over
the visible stellar hemisphere. \bz\ may be measured either
using a low or mid resolution spectropolarimeter such as FORS at the
European Southern Observatory ($R$ up to $\sim 2000$), ISIS on the
William Herschel Telescope ($R$ up to $\sim 8000$), or the Main Stellar
Spectrograph of the Special Astrophysical Observatory of the Russian
Academy of Sciences ($R$ up to 6000), that can measure the small
circular polarisation in the broad wings of H or He lines
\citep{Bagnetal02,Aznaetal04}, or using a high resolution
spectropolarimeter such as ESPaDOnS at the Canada-France-Hawaii
Telescope to measure the circular polarisation signal in the core of
the Balmer H$\alpha$ line \citep{Landetal15}. We have used all of
these spectropolarimeters in our searches for weak fields.

\subsection{Observations of WD\,2047+372 with ESPaDOnS}


ESPaDOnS is a high resolution cross-dispersed echelle
spectropolarimeter, constructed for the Canada-France-Hawaii
Telescope. The instrument produces an almost complete spectrum between
about 3800 and 10\,400\,\AA, with a typical resolving power of about
65\,000. ESPaDOnS can measure the intensity spectrum $I$, together
with one of the other three Stokes components $Q$, $U$ (linear
polarisation) or $V$ (circular polarisation) as functions of
wavelength. (The $I$ spectrum, of course, is simply a measurement
of the incoming stellar flux as a function of wavelength, as modified
by the terrestrial atmosphere, by the telescope and instrument
transmission functions, and by the efficiency of the detector.)

The instrument consists of a polarisation analyser mounted at the
Cassegrain focus of the 3.6-m telescope, and a bench mounted
spectrograph. In the polarisation analyser, a series of Fresnel rhombs
are combined in such a way as to act as a fully achromatic rotating
$\lambda/4$ or $\lambda/2$ waveplate. The retarder is followed by a
small-angle beam-splitting Wollaston prism, and the entrance aperture
is re-imaged onto two fibre optic cable ends\footnote{See
http://www.ast.obs-mip.fr/projets/espadons/espadons\_new/configs.html
for more details about the optical layout}.  The two beams which carry
the intensities of the two orthogonal polarisation states are
transferred by the fibre optic cables to a temperature-stabilised,
bench-mounted cross-dispersed echelle spectrograph which splits the
observed spectrum into 40 orders, ranging in wavelength extent per
order from about 100\,\AA\ (near 4000\,\AA) to more than 300\,\AA\
(near 10\,000\,\AA).

Because the entrances to the fibre optic cables follow the polarising
optics, and act as stops, the relative intensity of the two beams may
vary slightly with seeing and flexure between one exposure and the
next, even if the stellar beam is completely unpolarised.  As a result
the broad-band polarisation of the incoming beam cannot be measured
accurately with ESPaDOnS. Line polarisation may however be very
accurately measured with respect to the nearby continuum, which is in
fact set to zero by the ESPaDOnS pipeline. This situation is perfectly
satisfactory when the continuum may be assumed to be unpolarised.

The broad wings of the Balmer lines of white dwarfs cover wavelength
ranges comparable to the width of single orders of the echelle
spectrograph, and measuring the polarisation in their broad wings
presents almost the same problems as detecting polarisation in the
continuum. Practically, the polarisation in the outer Balmer line
wings of magnetic white dwarfs is largely undetectable by ESPaDOnS
(and presumably by other similar high resolution spectropolarimeters),
although (depending on the details of instrumental continuum
polarisation removal) some polarisation may still be detectable in the
inner Balmer line wings within a few \AA\ of the line core.
However, it is quite practical to measure the magnetic field from the
analysis of the sharp core of H$\alpha$.  Because this core is much
narrower than the FWHM of the full Balmer line, the $V/I$ polarisation
signal in the wings of the sharp core is much larger than that in the
broad wings, and is not affected by the removal of broad-band
instrumental polarisation. We have previously used high resolution
spectropolarimetry of this feature to set a very low upper limit on a
possible large-scale magnetic field in the bright white dwarf
40\,Eri\,B \citep{Landetal15}. In that study we showed that the line
core is indeed about as sensitive to non-zero \bz\ fields as we
expected, by measuring fields with the very similar H$\alpha$ line
cores found in several main sequence B stars. It is found that because
the line core is narrow and deep, this single feature makes possible
\bz\ field measurements that are competitive in terms of field
strength uncertainties to more conventional measurements using full
line wings.  Compared to low resolution spectropolarimeters, a further
major advantage of the use of ESPaDOnS is that its higher spectral
resolution allows us to simultaneously detect more subtle Zeeman
splitting in Stokes $I$. As a result, field measurements of white
dwarfs using high resolution provide excellent sensitivity and
accuracy for determination of the mean field modulus \bs\ of the
magnetic field.

\begin{figure}[ht]
\scalebox{0.35}{\includegraphics*{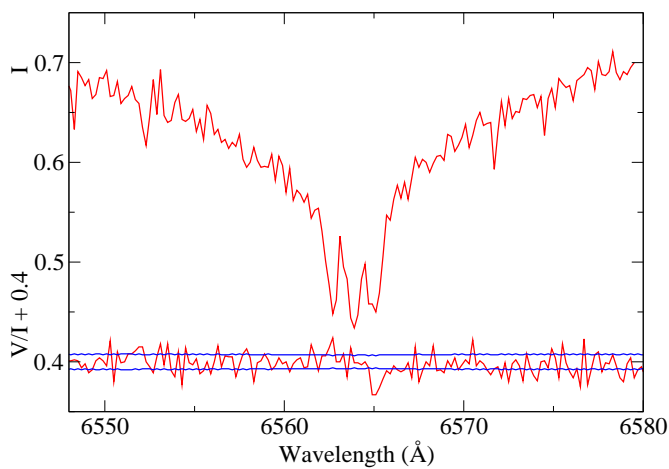}}
\caption{\label{Fig_esp_Halp_spec} Deep $I$ H$\alpha$ line core 
  (upper curve) and observed $V/I$ circular polarisation spectrum
  (lower curve) of \mwd, binned over intervals of 0.2\,\AA\ and
  arbitrarily normalised. The $V/I$ spectrum is moved
  upwards by +0.4 vertical units to facilitate comparison with the $I$
  spectrum. We also plot the $\pm 1\sigma$ level of photon noise
  bracketting the observed $V/I$ spectrum, both lines also shifted
  upwards by +0.4.}
\end{figure}

We acquired one ESPaDOnS circularly polarised spectrum of \mwd, on
2015 October 31 at 06:20\,UT, MJD 57326.255, requiring $4 \times 814 =
3256$\,s total shutter time. A small window around H$\alpha$ is shown
in Figure~\ref{Fig_esp_Halp_spec}.  The H$\alpha$ spectra from two
normalised overlapping orders have been combined by
weighted averaging. Then both the intensity ($I$) and circularly
polarised ($V/I$) spectra have been binned over wavelength intervals
of 0.2\,\AA, which substantially improves the S/N ratio of each
plotted point of the $I$ and $V/I$ Stokes components, without
significantly degrading the resolution of the features in these
spectra.

\begin{figure}[ht]
\scalebox{0.35}{\includegraphics*{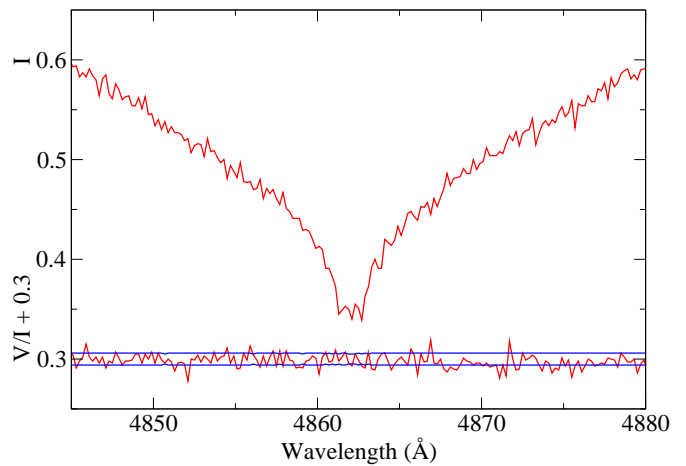}}
\caption{\label{Fig_esp_Hbet_spec} H$\beta$ line core of \mwd. 
  The meaning of symbols is as in Figure~\ref{Fig_esp_Halp_spec}. The
  $V/I$ signal and its associated $\pm 1 \sigma$ error bars have been
  shifted vertically by +0.3.  }
\end{figure}

The very obvious magnetic signature in Fig.~\ref{Fig_esp_Halp_spec} is
the clear splitting of the line core into a Zeeman
triplet. Strikingly, the two flanking $\sigma$ components are no wider
than the central $\pi$ component, suggesting that the dispersion in
local field strength $|B(x,y)|$ is not a great deal larger than about
0.2 or 0.3 of its actual value. The wavelength position of each
of the components of the Zeeman triplet was measured using the IRAF
``splot'' function to measure the barycentre of the component after
subtracting the linear continuum marked by the user.  The uncertainty
of these measurements was estimated by repeating them. The results
are that the separation of the blue $\sigma$ component from the $\pi$
component is about $1.152 \pm 0.01$\,\AA, and that of the red
component is about $1.139 \pm 0.01$\,\AA, for an average of $1.145 \pm
0.007$\,\AA. Using the simple expression for the separation of the
$\sigma$ components from the wavelength of the line in zero field,
\begin{equation}
\Delta \lambda_{\rm Z} = C_{\rm z} g_{\rm eff} \lambda_0^2 B,
\end{equation}
where $C_z = 4.67\,10^{-13}$\,\AA$^{-1}$\,G$^{-1}$ is a constant,
wavelengths are measured in \AA\ and the field in G, the Land\'{e}
factor of the H$\alpha$ line $g_{\rm eff} = 1.0$
\citep{CasLan94}, and the mean measured separation corresponds to a
mean magnetic field modulus, suitably averaged over the visible
hemisphere, of $\bs = 56.9 \pm 0.4$\,kG. It is quite clear that \mwd\
hosts a magnetic field of a little less than 60\,kG.

The core of the H$\beta$ line of \mwd, shown in Figure~\ref{Fig_esp_Hbet_spec}, has an abnormal flat bottom,
and if the weak structure in that core is interpreted as incipient
Zeeman splitting, the estimated field is fully compatible in strength
to that measured from the H$\alpha$ line core. 

Although the circular polarisation seen in Figure~\ref{Fig_esp_Halp_spec} is non-zero only
at a fairly marginal level, the circular polarisation $V/I$ shows weak
peaks of opposite sign at the positions of the two $\sigma$
components. This weak signal in $V/I$ is consistent with a slightly
non-zero value of \bz. The field can be evaluated using the standard
expression for measuring the separation between the mean wavelengths
of the right and left circularly polarised line cores,
\begin{equation}
\bz = -2.14\,10^{12} \frac{\int v V(v) {\rm d}v}
      {g_{\rm eff}\lambda_0 c \int(I_{\rm cont} - I(v)){\rm d}v} 
\end{equation}
where $I$ and $V$ are expresed as functions of velocity $v$ relative
to line centre at $\lambda_)$, and $c$ is the speed of light
\citep{Math89,Donaetal97}. The Land\'{e} factor 
$g_{\rm eff} = 1.0$ \citep{CasLan94}.  The numerical evaluation of
this expression and its uncertainty are discussed by
\citet{Landetal15}.

Evaluating the integrals over the full width and depth of the line
core below the broad line, we find $\bz = +6000 \pm 1670$\,G, barely
significant at the $3.6 \sigma$ level.

\subsection{Observations of WD\,2047+372 with ISIS at the WHT}
The Intermediate dispersion Spectrograph and Imaging System (ISIS) is
a medium resolution spectrograph, equipped with polarimetric optics,
mounted at the Cassegrain focus of the 4.2\,m William Herschel
Telescope (WHT) on La Palma.  The use of a dichroic filter permits
simultaneous observing in two arms, one optimised for the blue and one
for the red wavelength range.  In our observations we used the grism
600B in the blue arm, with a dispersion of 0.45\,\AA\ per pixel and
spectral resolving power of 2200 (1\arcsec slit width), and grism 1200R in
the red arm, with a dispersion of 0.26\,\AA\ per pixel and a spectral
resolving power of 7400 (1\arcsec\ slit width).  The grisms were mounted
such as to cover the spectral ranges of $\sim 3700-5200$\,\AA\ and
$\sim 6100-6850$\,\AA\ in the blue and in the red arm, respectively.

The polarimetric optics consist of an achromatic $\lambda/4$ retarder
waveplate followed by a Savart plate. A decker with three $18\arcsec$
strips prevent the overlapping of the two beams split by the Savart
plate, and allows sampling the sky background on two regions adjacent
to the target. The instrument can measure both continuum and line
polarisation. Its magnetographic capabilities are somewhat limited by
the fact that its spectral resolution is not sufficient to resolve
sharp metal lines; some flexure issues (typical of all Cassegrain
mounted instrument) may also generate spurious polarisation signals,
as described in detail by \citet{Bagetal13}. Nevertheless, ISIS has
proved a very efficient instrument for the detection of magnetic
fields in white dwarfs, and its spectral resolution may be sufficient
to resolve Zeeman splitting of H$\alpha$ for field strength $\ga
80$\,kG. CCD readout was set to the 'FAST' mode, with no pixel
binning, but only a window of 405 columns and 4200 rows, centred about
the target and background spectra, was read out, to minimise
overheads.

Observations of \mwd\ were carried out on the night from 31 Aug
2015 to 01 Sep 2015 (mid exposure at 01:58 UT on 2015-09-01, or MJD
57266.082), as a part of an 8\,night long magnetic survey of bright
white dwarfs. Total exposure time was 3360\,s, split into eight
exposures with position angles of the retarder waveplate at 315, 45,
45, 315, 315, 45, 45, 315\degr. The target is relatively faint and the
reason for using eight exposure instead of four was not dictated by
the risk of CCD saturation. We decided to split the observations in a
larger number of exposures than the minimum strictly required to check
for magnetic field variation due to rapid rotation. We note that
overheads for readout and retarder waveplate rotation were actually
short, of the order of 15\,s. Wavelength calibration was obtained
after the last science frame, keeping the instrument and the telescope
at the same position reached at the end of the exposure series. Over
the course of the science exposure the telescope zenith angle changed
by $\sim 10\degr$ and the instrument rotated by $\sim 10\degr$. 

\begin{figure}
\scalebox{0.45}{\includegraphics*[trim={0.1cm 5.5cm 0.3cm 3cm},clip]{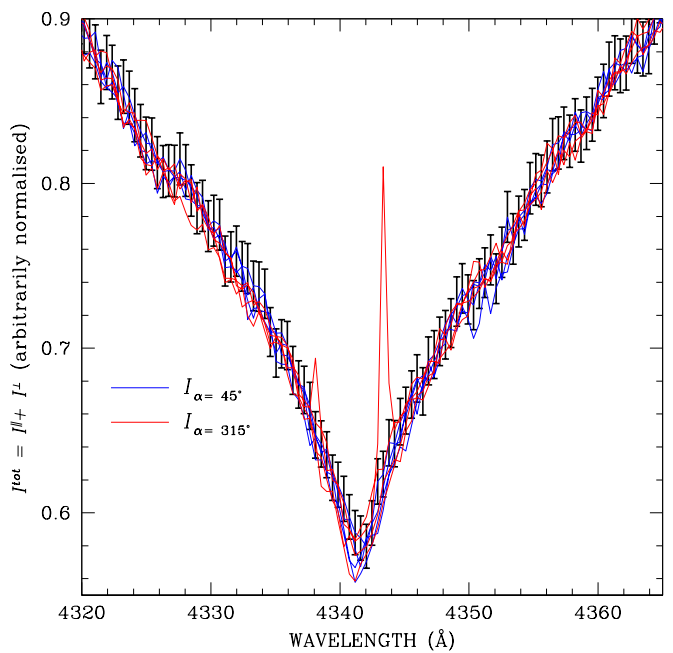}}
\caption{\label{Fig_Flexures} \mwd, H$\gamma$ Balmer line:
  the sum of fluxes in the
  parallel and perpendicular beam split by the Savart plate,
  normalised to a pseudo-continuum, for all eight exposures
  obtained in the observing series, overplotted with the error bars.
}
\end{figure}

The impact of possible significant instrument flexure on the spectral
lines of \mwd\ was checked using the method outlined by
\citet{Bagetal13}, who suggested that overlapping the normalised
profiles of spectral lines obtained in successive expoures could
reveal harmful flexure.  Figure~\ref{Fig_Flexures} shows the sum of
the fluxes in the two beams split by the Savart plate measured during
the various exposures, compared to the error bars due to photon-noise.
The sum of the two fluxes should be unpolarised, and the offsets
between fluxes obtained at different positions of the retarder
waveplate that exceed photon noise error bars should be entirely
ascribed to instrument flexures or other non-statistical noise. One
can see that with the exception of one profile affected by a cosmic ray,
all fluxes (after re-normalisation) are consistent with photon-noise
error bars.  We conclude that the impact of flexures on the broad H
Balmer lines was nearly negligible, although some 3 ot $5 \sigma$
detections in the null fields suggest that error bars may be slightly
underestimated. \citep[For a thorough discussion of the role of the
null field as a quality check parameter, see][]{Bagnetal12}

\begin{figure}
\scalebox{0.45}{\includegraphics*[trim={0.8cm 4.8cm 0.1cm 3.7cm},clip]{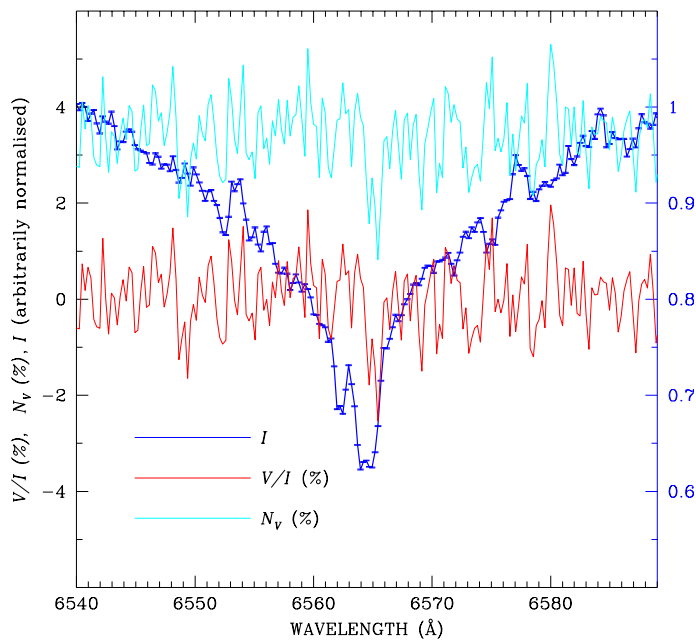}}
\caption{\label{Fig_Halpha_ISIS} H$\alpha$ Stokes $I$ (blue solid line) and $V/I$ (red
  solid line) profiles of \mwd\ obtained with ISIS. The null profile
  is plotted with a solid light blue line, offset by $+3.5$\,\% for
  display purpose, and the left scale describes these data. The scale
  for the $I$ spectrum is on the right.  }
\end{figure}

Remarkably, it appears that the resolving power of the red measurement
of the spectrum of \mwd\ with ISIS is high enough that the splitting
of the core of H$\alpha$ in a 57\,kG field is detectable. In
Figure~\ref{Fig_Halpha_ISIS} we show the $I$ and $V/I$ spectra
observed with ISIS.  The ISIS spectrum is very similar to the ESPaDOnS
spectrum, and one can clearly see Zeeman splitting in the
H$\alpha$ line core of the ISIS spectrum. In addition, although the
$V/I$ signal is not significantly different from zero, it appears that there
may be a small polarisation signal in the $\sigma$ components of the
line core that is similar to that observed by ESPaDOnS. It is found 
that the field components \bs\ and \bz\ observed by ISIS are
not very different from those observed with ESPaDOnS; the pairs of
observation are in good enough agreement that they do not make a
strong case for variation between the two observations about two
months apart, except for what does appear to be a significant change in
the shape of the line core components.


\begin{figure}
\scalebox{0.45}{\includegraphics*[trim={0.8cm 7.5cm 0.1cm 3cm},clip]{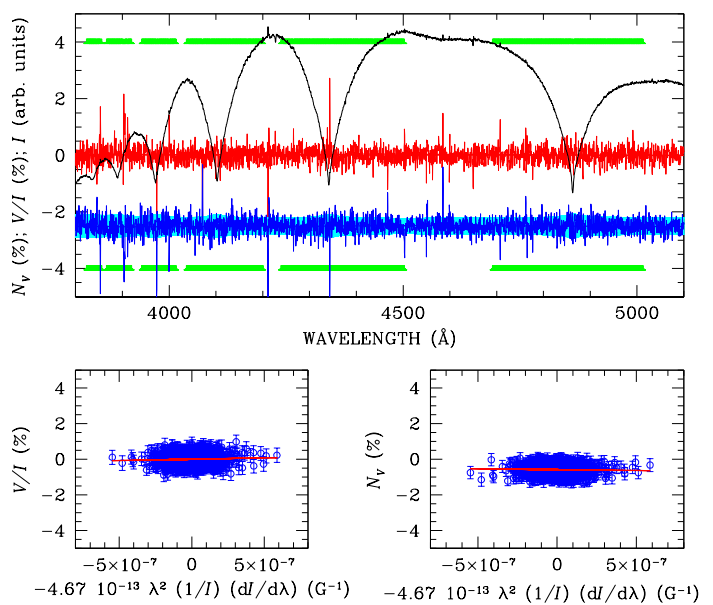}}
\caption{\label{Fig_WD2047_ISIS_Blue} Spectropolarimetry of \mwd.
In the upper panel, the black solid line shows the intensity profile,
the shape of which is heavily affected by the transmission function of
the atmosphere + telescope optics + instrument. The zero level of
intensity is very close to the bottom of the panel, at $-5$ on the
left-hand scale. The red solid line is the reduced Stokes $V/I$
profile in \% units. Photon-noise error bars are centred around
$-2.5$\,\% and appear as a light blue background, to which the null
profile (also offset by $-2.5$\,\% for display purpose) is superposed
(blue solid line). The two bottom panels show the best-fit obtained by
fitting with a straight line $V/I$ as a function of the quantity
$\propto \lambda^2 (1/I)\ \mathrm{d}I/\mathrm{d}\lambda)$. The green
bars at the top and bottom of the upper panel show the spectral
regions included in the correlation plots in the small lower panels. }
\end{figure}

The field strength \bz\ was determined using the weak-field
approximation, 
\begin{equation}
V/I = -g_{\rm eff} C_{\rm z} \lambda^2 \frac{1}{I} 
       \frac{{\rm d}I}{{\rm d}\lambda} \bz, 
\end{equation}
as the slope of the correlation line of the pixel-by-pixel circular
polarisation $V/I$ as a function of the local value of $-g_{\rm eff}
C_z \lambda^2 (1/I)({\rm d}I/{\rm d}\lambda)$, a method discussed in detail by
\citet{Bagnetal02,Bagnetal12}. As in Eq.~(1), $g_{\rm eff} = 1$
is the effective Land\'{e} factor of the Balmer lines, $g_{\rm
eff} = 1.25$ is assumed for the metal lines \citep{Bagnetal12}, and
$\lambda$ is the wavelength in \AA\ units. Using this technique we
measured the magnetic field from all H Balmer lines from H$\alpha$ to
H9. As a quality check we also measured the magnetic fields from the
null profiles.  The results are presented in Table~\ref{Tab_ISIS}.
Figure~\ref{Fig_WD2047_ISIS_Blue} shows our observations and the
results of the least-square technique used for field determination on
the H Balmer lines from H$\beta$ down to H$9$.

\begin{figure}
\scalebox{0.45}{\includegraphics*[trim={0.8cm 4.5cm 0.1cm 2.8cm},clip]{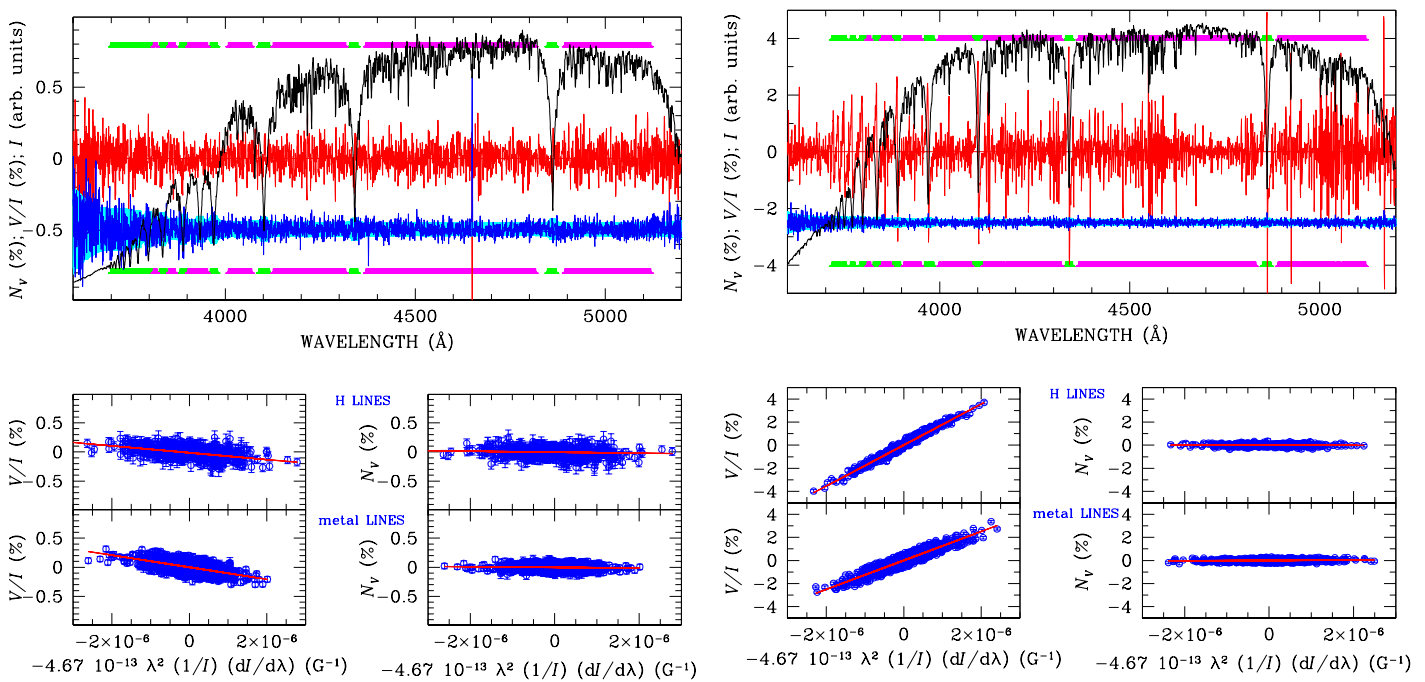}}
\caption{\label{Fig_HD201601} Same as Fig.~\ref{Fig_WD2047_ISIS_Blue}
  but for the magnetic Ap star $\gamma$\,Equ. In the top panel, the magenta
  lines highlight the regions covered by metal lines that have been used for
  a further estimate of the magnetic field. 
}
\end{figure}
\begin{figure}
\scalebox{0.45}{\includegraphics*[trim={0.8cm 4.5cm 0.1cm 3.0cm},clip]{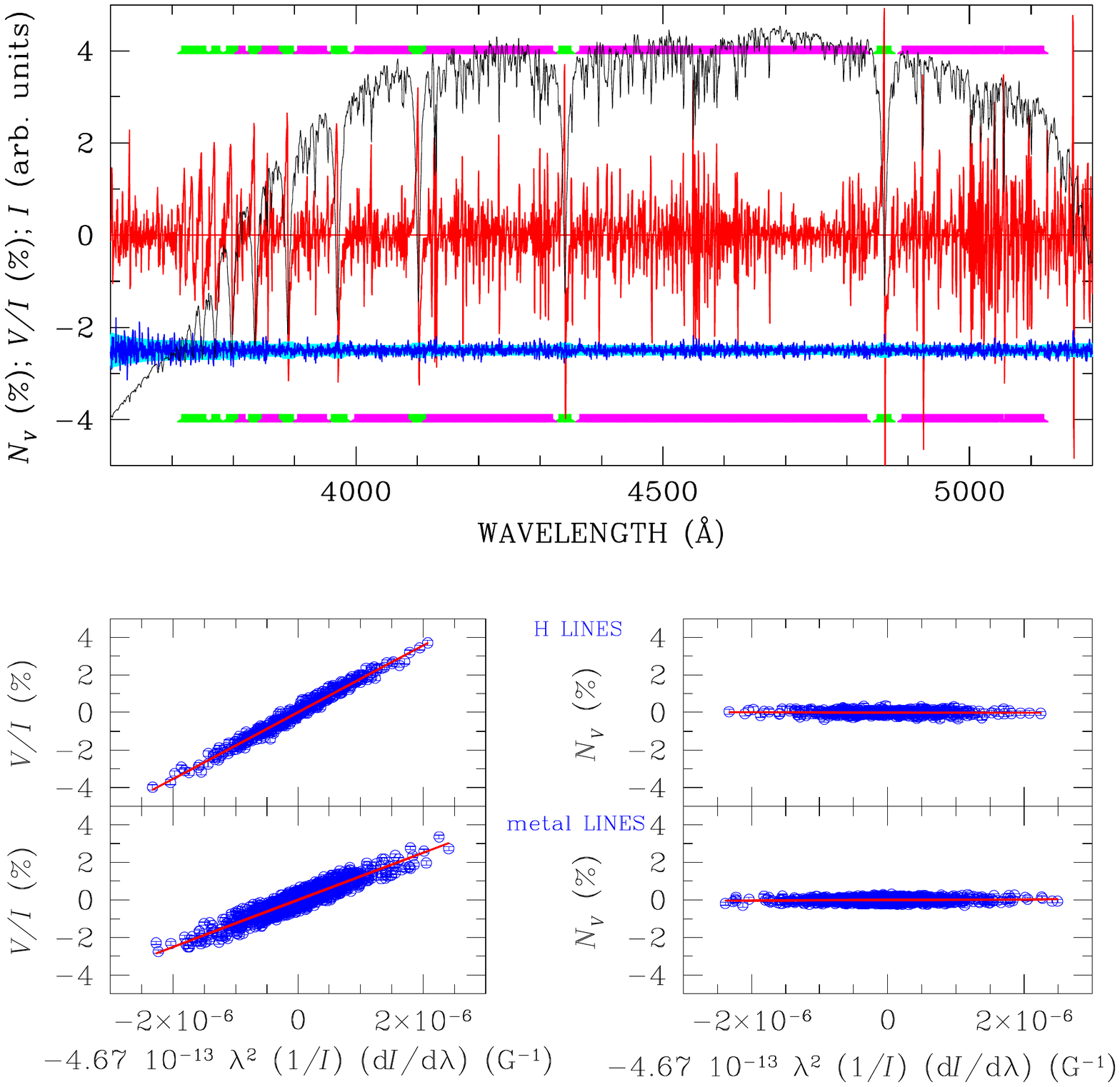}}
\caption{\label{Fig_HD215441} Same as Fig.~\ref{Fig_WD2047_ISIS_Blue}
  but for the magnetic Ap star HD\,215441 (Babcock's star).
}
\end{figure}

\begin{table*}
  \caption{\label{Tab_ISIS} Magnetic field measurements for three stars
    obtained with the ISIS instrument. }
  \begin{center}
    \begin{tabular}{lr@{$\pm$}lr@{$\pm$}l|r@{$\pm$}lr@{$\pm$}l|r@{$\pm$}lr@{$\pm$}l}
      \hline\hline
      \multicolumn{13}{c}{}\\
   Star:  &
     \multicolumn{4}{c}{WD\,2047+372} &
     \multicolumn{4}{c}{HD\,215441}   &
     \multicolumn{4}{c}{HD\,201601}  \\
   Epoch: &
         \multicolumn{4}{c}{2015-09-01 01:58 UT} &
         \multicolumn{4}{c}{2015-08-29 05:56 UT} &
         \multicolumn{4}{c}{2015-08-31 03:06 UT}\\
Exp. time, peak S/N: &
         \multicolumn{4}{c}{3360\,s,\ \  576\, per \AA} &
         \multicolumn{4}{c}{ 960\,s,\ \ 1740\, per \AA} &
         \multicolumn{4}{c}{ 120\,s,\ \ 3770\, per \AA} \\ [2mm]
                  &\multicolumn{2}{c}{\bz\ (G)} &\multicolumn{2}{c}{\nz\ (G)} &
                   \multicolumn{2}{c}{\bz\ (G)} &\multicolumn{2}{c}{\nz\ (G)} &
                   \multicolumn{2}{c}{\bz\ (G)} &\multicolumn{2}{c}{\nz\ (G)} \\ [2mm]
\hline
    H$\alpha$ &$ 790$&  430&$  910$&  400&$14160$&  71&$   50$&  77&$ -525$&  94&$   39$&  32\\
    H$\beta$  &$ 940$& 1000&$  185$&  960&$16525$& 300&$   45$& 160&$ -680$&  67&$  -36$&  57 \\
    H$\gamma$ &$ 490$& 1080&$ -730$& 1200&$17550$& 230&$  -20$& 170&$ -700$&  70&$  106$&  45 \\
 H$\delta$--H9&$ +40$& 1070&$  -75$& 1130&$18050$& 140&$  -55$&  85&$ -505$&  48&$ -147$&  38 \\
 All H lines  &$ 610$&  240&$ -715$&  240&$17785$& 115&$  -50$&  65&$ -520$&  33&$  -44$&  19 \\
 blue metal lines
              & \multicolumn{2}{c}{}& \multicolumn{2}{c}{} &$12510$& 100&$  130$&  35&$-1036$&  30&$  -50$&  15 \\
 red  metal lines
              & \multicolumn{2}{c}{}& \multicolumn{2}{c}{} &$ 5090$& 170&$   50$&  25&$-1050$&  20&$   50$&  10 \\
 \hline
    \end{tabular}
\tablefoot{``Blue metal lines'' and ``red metal lines'' are all 
spectral lines in the regions not occupied by H Balmer lines, detected
in the blue and red arm, respectively.  }
\end{center}
  \end{table*}

\subsubsection{Other observations with ISIS}
The ISIS instrument has not been extensively used as a
magnetograph. To the best of our knowledge, the only ISIS field measurements
reported in the literature are those by \citet{Leone07}, who described
a measurement of four magnetic Ap stars, and the paper by
\citet{Landetal15}, who reported observations of the white dwarf
40\,Eri\,B. Therefore, in the course of our observing campaign, we
have observed a number of well known magnetic stars and
carried out a few experiments which will be discussed in detail in a
forthcoming paper. Here we limit ourselves to discussing some
observations obtained during the same run as \mwd, which support the
reliability of its field measurement.

During our run we observed the well known magnetic star
HD\,215441 (Babcock's star), and the long period ($P \sim
100$\,years!)  magnetic star HD\,201601 = $\gamma$\,Equ. The results
of our measurements are given in Table~\ref{Tab_ISIS}. They are
globally consistent with previous literature data.

Of special interest is the case of HD\,201601, for which the average
of field measurements made using the metal line spectrum (i.e.
excluding windows around each Balmer line from the analysis) is close
to $\bz \approx -1000$\,G. The ensemble of more than 60\,yr of \bz\
data available for the star, with many new measurements, has recently
been studied by \citet{Bychetal06,Bychetal16}, who fit a sine wave
variation to all the available \bz\ values, most of which are based on
analysis of spectropolarimetry of metal line spectra. They derive a
rotation period of $97.16 \pm 3.15$\,yr, with zero point at the
negative extremum that occurred on about JD\,241\,7795.0. (Note that
in the ephemeris as described in Eqs (2) and (3) of
\citet{Bychetal16}, the argument $\phi$ of the sine function in Eq.
(2) should be replaced by $\phi - \pi/2$ in order for the new results
to agree with the 2006 results.)

When their (corrected) new ephemeris and fitted sine wave parameters
are used to compute the expected \bz\ value for our measurement, their
ephemeris predicts $\bz \approx -740$\,G.  This is (probably)
significantly discrepant with our measurement of $-1000$\,G.  Since
even the huge data set used by \citet{Bychetal16} does not yet cover
one full rotation cycle, the period that they find is certainly
affected by scatter due to different instrumental systems of
measurement of \bz\ \citep[see][]{Landetal14}, which means that the
uncertainty in the period that they derive, $\pm 3.15$\,yr, is probably
underestimated.  Our measurement suggests that the true rotation
period may be a few years, perhaps even a decade, longer than the
period of \citet{Bychetal16}. However, the discrepancy could also
arise because our instrumental measurement system for \bz, even
restricting ourselves to the value from the blue arm of ISIS, is
significantly different from many of the other instrumental systems
used to measure \bz\ in the past.

Note that the average of our Balmer line measurements of the field of
$\gamma$\,Equ, about $-520$\,G, is consistent with the observation by
\citet{Bychetal16} that the values of \bz\ measured for HD\,201601
using Balmer lines tend to have smaller amplitude than those measured
using metal lines, and in fact our measurement is in good agreement
with the recent \bz\ measurements of HD\,201601 from dimaPol shown in
Figure~6 of \citet{Bychetal16}.

Overall, our measurement of \bz\ for $\gamma$\,Equ confirms that
field measurements made with ISIS have the expected sensitivity and
accuracy, and are consistent with measurements from other
well-tested spectropolarimeters. 

The observations of HD\,215441 reveal a somewhat more complex result.
Table~\ref{Tab_ISIS} shows a general concordance among the Balmer line
field values for HD\,215441, but also a modest systematic increase in
the detected field from H$\alpha$ to the upper Balmer lines. This
systematic trend probably arises from the fact that different Balmer
lines have different darkening and line weakening rates towards the
limb, and thus weight the local $B_z(x,y)$ field differently from one
another in the average over the visible hemisphere.  However, the
observed field strength is within the range of \bz\ values found
(using the wings of the Balmer H$\beta$) for HD\,215441 by
\citet{BorrLand78}. It is clear that the weak-field approximation of
Eq.\,(3) used to measure \bz\ from the ISIS Balmer line spectrum is
still valid for fields up to at least $\bz \sim 20$\,kG. Thus we are
confident that Eq.\,(3) is also valid for the (weaker) \bz\ field
measured in \mwd, using still broader Balmer lines than those of
HD\,215441.

In contrast, there is a large discrepancy between \bz\ field
measurements of HD\,215441 obtained using metallic lines in the red
and in the blue arm, and a smaller discrepancy between the blue arm
metallic line \bz\ and the blue arm Balmer line \bz.  From the
instrumental point of view, one could suspect that the retarder
waveplate is affected by a chromatic problem, i.e., that the
retardation is optimised for the blue region, and departs
substantially from its nominal $\pi/4$ value at $\ga 6000$\,\AA.
However, the consistency of the very precise red and blue arm metal
line field values measured for $\gamma$\,Equ suggests instead that the
explanation for the observed discrepancy may be specific to the
observed star rather than due to instrumental reasons.  Babcock's star
has a very strong magnetic field, and the Zeeman splitting is totally
resolved in many spectral lines, with $\pi - \sigma$ separations of
more than 0.5\,\AA\ in some blue lines \citep{Landetal89}. In a
spectrum with resolved Zeeman patterns, the peaks in $V$ no longer
coincide with regions of large slope ${\rm d}I/{\rm d}\lambda$ in the
line profile, but instead are largest where the two sigma components
are strongest. Thus Eq.\,(3) ceases to be a useful approximation.

In a low resolution spectrum, instrumental broadening can recreate the
weak-field situation of Eq.\,(3) by blending the Zeeman components
together so that the (smoothed) peaks in $V$ still do coincide roughly with
(smoothed) line wings in the $I$ spectrum.  In the ISIS spectrum, the
resolving power is larger in the red arm than in the blue arm, by a
factor of about three. Furthermore, the typical Zeeman splitting,
proportional to $\lambda^2$ (see Eq\,(1) is about twice as large in the
red as in the blue arm spectra. As a result, the blue arm spectrum is
still mostly in the weak-field regime with peaks in $V$ coinciding
with line wings in $I$, while in the red arm spectrum this
approximation has begun to break down seriously: many peaks in $V$ are
more nearly coinciding with dips in $I$ than points of large slope
${\rm d}I/{\rm d}\lambda$. Thus the value of \bz\ derived with the
weak-field approximation with the blue arm spectrum is only moderately
smaller than values found using Balmer lines (which are certainly
still in the weak-field regime), while \bz\ from the red arm falls
well below the expected value.

The correctness of this argument is supported by analysis of a single
high resolution ESPaDOnS spectrum; in that spectrum the \bz\ field
strengths of about 15\,kG derived from various parts of the spectrum
using Eq.\,(2), and thus without using the weak-field
approximation of Eq.\,(3), are all consistent.

\section{The magnetic field of WD\,2047+372}

In the ESPaDOnS $I$ spectrum we see well-defined $\sigma$ components
that are not significantly broader than the central $\pi$ component,
quite unlike the situation of some other weak-field magnetic stars
such as WD\,2359$-434$, whose $\sigma$ components are clearly strongly
broadened \citep[cf.][]{Landetal12}. The narrowness of the $\sigma$
components in \mwd\ strongly suggests that the field at the time of
the \esp\ observation is relatively uniform in field modulus \bs, and
it may well be rather orderly. These data are consistent with a very
simple field geometry, one which could be approximated by a uniform
field in the hemisphere viewed at the time of the \esp\ observation.

Both measurements of \bz\ give values of about $0.1 \bs$ or less. If
we suppose that the field is approximately described by the oblique
rotator model (a roughly axisymmetric field fixed in the star, with
the magnetic axis inclined at an angle $\beta$ to the stellar rotation
axis, which in turn is inclined by the angle $i$ to the line of
sight), then the small value of \bz\ relative to \bs\ suggests that
the line of sight is directed approximately towards the stellar
magnetic equator, which would imply that at least one of $i$ or
$\beta$ is large. However, further field measurements are required to
establish whether this model is plausible or not.

\begin{figure}
\scalebox{0.35}{\includegraphics*[trim={0.1cm 0.5cm 0.3cm 3cm},clip]{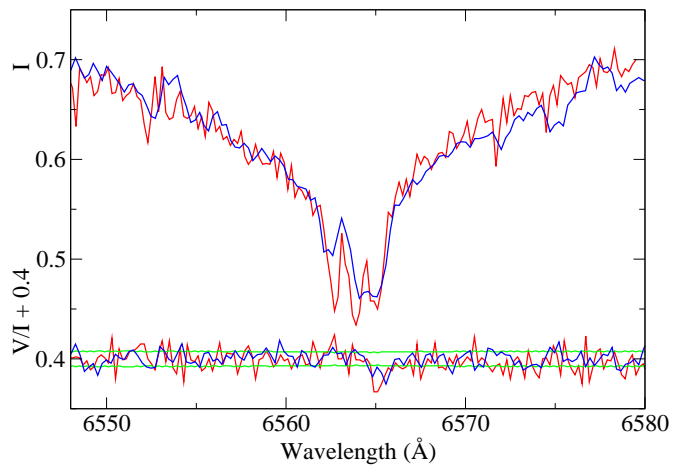}}
\caption{\label{Fig_Esp_ISIS_comparison} Comparison of the $I$ and
$V/I$ spectra of the core of H$\alpha$ as observed with ISIS on
2015-09-01 (blue solid line), and as observed with \esp\ on 2015-10-31
(red solid line). Both spectra have been normalised to 0.65 at 6555~\AA. The
$V/I$ spectra have been shifted upwards by $+0.4$ for ease of
comparison with the $I$ spectra.   }
\end{figure}

Because we have two spectra of \mwd, we can obtain some
information about possible variability of the magnetic field. A
comparison of the two H$\alpha$ $I$ spectra is shown in
Figure~\ref{Fig_Esp_ISIS_comparison}. The \esp\ spectrum here is
binned to pixels of 0.25~\AA\ to match the pixel spacing of the ISIS
spectrum. The two Zeeman split line cores are similar in width, but
the \bs\ field seems to have been slightly stronger at the time that
the ISIS spectrum was obtained. More striking is the obvious
difference in shape between the two line cores. It seems clear that
the observed field strength or structure of \mwd\ is somewhat variable
as the star rotates. The asymmetric form of the ISIS H$\alpha$ line
core may be the result of a combination of a somewhat non-uniform
field over the hemisphere viewed at the time of that observation,
combined with the Doppler shifts associated with significant rotation
velocity.

\section{Discussion and Conclusions}

The significance of the detection of this very weak field (``weak'' for a
white dwarf!) is that there are only two white dwarfs in which the
presence of a still weaker magnetic field is clearly established by
spectropolarimetry and observation of the H$\alpha$ line core
broadening: WD\,0446$-789$, with $\bs \sim 20$\,kG, and
WD\,2105$-820$, with $\bs \approx 43$\,kG
\citep{Koesetal98,Aznaetal04,Landetal12}. It has been suggested
that a very weak field is present in WD\,1105$-048$
\citep{Aznaetal04}, but our unpublished measurements have not yet
clearly confirmed, or rejected, this result.

If we look at the current sample of white dwarfs for which fields have
been firmly detected, and that show values of \bs\ that are always
below 120\,kG, which we list in Table~\ref{Tab_wk_fld_mwds}, we can
add only WD\,1653$+385$, with $\bs = 70$\,kG; WD\,0257$+080$, with
$\bs \approx 90$\,kG; WD\,2359$-434$, with $\bs\approx 110$\,kG, and
WD\,0322$-019$, with $\bs = 120$\,kG
\citep{Bergetal01,Aznaetal04,Koesetal09,Gianetal11,Zucketal11,Farietal11}.
Thus \mwd\ is a valuable addition to the very small sample of the
weakest known white dwarf magnetic fields. Furthermore, of these seven
stars, even the most elementary magnetic and geometric model has been
proposed for only one, WD\,2105-820 \citep{Landetal12}.

\begin{table}[ht]
\begin{center}
\caption{White dwarfs with confirmed fields $\bs \le 120$\,kG}
\label{Tab_wk_fld_mwds}
\begin{tabular}{lrrrccc }\hline 
   Star    &  \te       &\bz\ range     &  \bs\  \\
           &  (K)       &      (kG)     &     (kG)    \\
\hline
WD\,0446$-789$ &  24440     &--2.5 to --5.5 & $\sim 20$ \\
WD\,2105$-820$ &  10600     & $-8$ to $-11$ & 43        \\
WD\,2047$+372$ &  14710     & $+2$ to $+6$  & 57        \\
WD\,1653$+385$ &  5900      &               & 70        \\
WD\,0257$+080$ &  6430      & $\sim 35$     & 90        \\
WD\,2359$-434$ &  8650      & 3             & 110       \\
WD\,0322$-019$ &  5310      &               & 120       \\
\hline
\end{tabular}
\tablefoot{The data in the table are from the references cited in
this section of the text.}
\end{center}
\end{table}

Summary lists of white dwarfs with weak magnetic fields have recently
been published by \citet{KawkVenn12,Ferretal15}. The most recent list,
from \citet{Ferretal15}, contains 15 white dwarfs which are supposed to
have magnetic fields below about 120~kG. Our list in
Table~\ref{Tab_wk_fld_mwds} contains only seven stars.  This is due to
changes to the two recent lists, and to omission from our list of doubtful
detections, as detailed below.
\begin{itemize}
\item Magnetic field detections in NLTT 347 = WD 0005-048,
  G234$-4$ = WD\,0728+642, LB\,8827 = WD\,0853+163, LTT 4099 =
  WD\,1105$-048$, WD\,1531$-022$, G\,226-29 = WD\,1647+591, and
  WD\,2039$-682$ are in fact currently unconfirmed, although some of
  these stars may eventually be shown to have some of the smallest
  white dwarf fields.
\item 40\,Eri\,B has been shown to be {em non}-magnetic at the level of
  about 250\,G \citep{Landetal15}.
\item The \bs\ value of LP\,907$-37$ = WD\,1350$-090$ was not measured
  by \citet{Koesetal98}, but it is found that $\bs = 460 \pm 20$\,kG from
  their spectrum.
\item The field attributed to G\,227-28 = WD\,1820+609 is due to a
  typographical error in \citet{Putn97}, who incorrectly recorded a
  non-detection measured by \citet{LiebStoc80}. 
\item WD\,2329$-291$ is not a white dwarf but an sdB star
  \citep{Gianetal11}.
\item The \bs\ field of LTT 9857 = WD\,2359$-434$ is not 9.8\,kG but
  slightly exceeds 100\,kG \citep{Koesetal09}.
\end{itemize}
It seems very probable that the small number of very weak magnetic
fields now known in white dwarfs is due to reaching the current
practical detection threshold, as detecting fields so weak rapidly
becomes increasingly difficult as one looks at fainter and fainter
stars. In fact, four of the seven weak-field magnetic white dwarfs
listed in Table~\ref{Tab_wk_fld_mwds} are brighter than $V = 13.5$,
and thus are among the brightest known magnetic white dwarfs.

\begin{figure}[ht]
\scalebox{0.35}{\includegraphics*{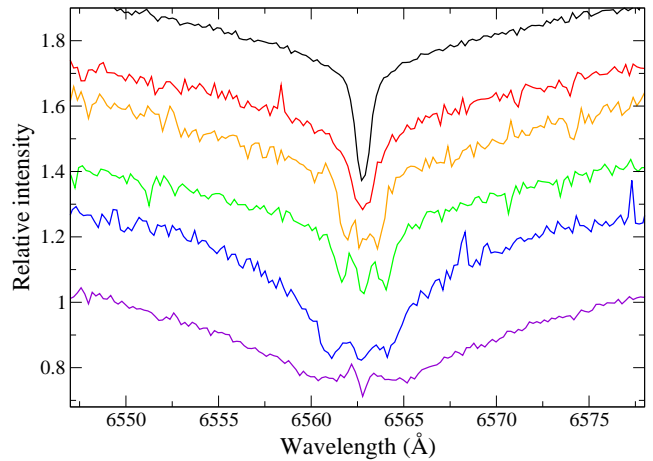}}
\caption{\label{Fig_comp_low_bz} Comparison of the intensity
  spectra of a non-magnetic white dwarf (WD\,2007$-303$, top) followed
  in order of increasing \bs\ by the DA stars WD\,0446$-789$,
  WD\,2105$-820$, \mwd, WD\,0257+080, and WD\,2359$-434$. All
  spectra have been rebinned in bins of width 0.2\,\AA\ (in some cases
  two spectra have been averaged) to improve S/N in the plot,
  normalised to 1.0 at 6550 and 6575\,\AA, Doppler shifted to the rest
  frame of H$\alpha$, and then offset vertically for clarity.  }
\end{figure}

We have available intensity spectra of five of the seven magnetic
white dwarfs of Table~\ref{Tab_wk_fld_mwds} (the DA stars), mostly
observed for the SPY project \citep{Koesetal98,Koesetal09} and
obtained from the ESO Archive.  The appearance of H$\alpha$ in these
stars is shown in Figure~\ref{Fig_comp_low_bz}. It will be noticed
that the rather well-defined Zeeman triplet of \mwd\ is unusual (and
as mentioned above probably reflects a field geometry such that $|B|$
varies by only perhaps 20 or 30\,\% across the surface of the
visible hemisphere). The weaker fields of WD\,0446$-789$ and
WD\,2105$-820$ do not show clear resolution into components because
the splitting is not large enough relative to the intrinsic width of
the Zeeman components, while the larger fields of WD\,0257+080 and
WD\,2359$-434$ are apparently rather non-uniform over the visible
hemisphere, thus broadening the $\sigma$ (but not the $\pi$)
components.

Our short list of five magnetic white dwarfs with \bs\ below 100\,kG,
of which three are brighter than $V = 13.5$, makes it possible for us
to estimate very roughly the fraction of white dwarfs in a
magnitude-limited sample that have extremely weak magnetic fields.
When we search the list of all white dwarfs in the Vizier catalogue of
McCook \&
Sion\footnote{http://www.astronomy.villanova.edu/WDCatalog/index.html},
we find 67 DA white dwarfs. We have not yet searched all 60+ white
dwarfs with the best currently possible uncertainties (nor has anyone
else), but it already appears (with rather weak statistical
significance) that at least about 3--4\% of DA white dwarfs probably
possess a very weak magnetic field, of the order of 10--100\,kG, a
situation previously suggested by \citet{SchmSmit95}.

The discovery of the very weak field of \mwd\ confirms that it is
practical to detect and study magnetic fields in white dwarfs having
\bz\ of the order of a few kG and $\bs \ga 20$\,kG with several
currently available facility instruments, down to a limiting magnitude
of the order of $V \sim 14 - 15$. We should thus soon have a
well-observed sample of at least several stars at the presently
observable low end of the distribution of white dwarf magnetic field
strength which can be studied in detail for clues that they may yield
about the origin and evolution of these fields.

\begin{acknowledgements}

  We thank the referee, Dr Stefan Jordan, for his careful reading and
  valuable comments.

  JDL acknowledges financial support from the Natural Sciences and
  Engineering Research Council of Canada. GV acknowledges the Russian
  Foundation for Basic Research (RFBR grant N15-02-05183).

  This research has made use of the VizieR catalogue access tool, CDS,
  Strasbourg, France. The original description of the VizieR service
  was published in A\&AS 143, 23.

  Based in part on data products from observations made with ESO
  Telescopes at the La Silla Paranal Observatory under programme IDs
  165.H-0588 and 167.D0407, and provided by the ESO Archive. 

\end{acknowledgements}

\bibliography{MyBiblio}

\begin{thebibliography}{39}
\expandafter\ifx\csname natexlab\endcsname\relax\def\natexlab#1{#1}\fi

\bibitem[{{Alecian} {et~al.}(2013){Alecian}, {Wade}, {Catala}, {Grunhut},
  {Landstreet}, {Bagnulo}, {B{\"o}hm}, {Folsom}, {Marsden}, \&
  {Waite}}]{Alecetal13}
{Alecian}, E., {Wade}, G.~A., {Catala}, C., {et~al.} 2013, \mnras, 429, 1001

\bibitem[{{Auri{\`e}re} {et~al.}(2015){Auri{\`e}re}, {Konstantinova-Antova},
  {Charbonnel}, {Wade}, {Tsvetkova}, {Petit}, {Dintrans}, {Drake}, {Decressin},
  {Lagarde}, {Donati}, {Roudier}, {Ligni{\`e}res}, {Schr{\"o}der},
  {Landstreet}, {L{\`e}bre}, {Weiss}, \& {Zahn}}]{Aurietal15}
{Auri{\`e}re}, M., {Konstantinova-Antova}, R., {Charbonnel}, C., {et~al.} 2015,
  \aap, 574, A90

\bibitem[{{Aznar Cuadrado} {et~al.}(2004){Aznar Cuadrado}, {Jordan},
  {Napiwotzki}, {Schmid}, {Solanki}, \& {Mathys}}]{Aznaetal04}
{Aznar Cuadrado}, R., {Jordan}, S., {Napiwotzki}, R., {et~al.} 2004, \aap, 423,
  1081

\bibitem[{{Bagnulo} {et~al.}(2013){Bagnulo}, {Fossati}, {Kochukhov}, \&
  {Landstreet}}]{Bagetal13}
{Bagnulo}, S., {Fossati}, L., {Kochukhov}, O., \& {Landstreet}, J.~D. 2013,
  \aap, 559, A103

\bibitem[{{Bagnulo} {et~al.}(2012){Bagnulo}, {Landstreet}, {Fossati}, \&
  {Kochukhov}}]{Bagnetal12}
{Bagnulo}, S., {Landstreet}, J.~D., {Fossati}, L., \& {Kochukhov}, O. 2012,
  \aap, 538, A129

\bibitem[{{Bagnulo} {et~al.}(2002){Bagnulo}, {Szeifert}, {Wade}, {Landstreet},
  \& {Mathys}}]{Bagnetal02}
{Bagnulo}, S., {Szeifert}, T., {Wade}, G.~A., {Landstreet}, J.~D., \& {Mathys},
  G. 2002, \aap, 389, 191

\bibitem[{{Bergeron} {et~al.}(2001){Bergeron}, {Leggett}, \&
  {Ruiz}}]{Bergetal01}
{Bergeron}, P., {Leggett}, S.~K., \& {Ruiz}, M.~T. 2001, \apjs, 133, 413

\bibitem[{{Borra} \& {Landstreet}(1978)}]{BorrLand78}
{Borra}, E.~F. \& {Landstreet}, J.~D. 1978, \apj, 222, 226

\bibitem[{{Bychkov} {et~al.}(2006){Bychkov}, {Bychkova}, \&
  {Madej}}]{Bychetal06}
{Bychkov}, V.~D., {Bychkova}, L.~V., \& {Madej}, J. 2006, \mnras, 365, 585

\bibitem[{{Bychkov} {et~al.}(2016){Bychkov}, {Bychkova}, \&
  {Madej}}]{Bychetal16}
{Bychkov}, V.~D., {Bychkova}, L.~V., \& {Madej}, J. 2016, \mnras, 455, 2567

\bibitem[{{Casini} \& {Landi Degl'Innocenti}(1994)}]{CasLan94}
{Casini}, R. \& {Landi Degl'Innocenti}, E. 1994, \aap, 291, 668

\bibitem[{{Donati} \& {Landstreet}(2009)}]{DonaLand09}
{Donati}, J.-F. \& {Landstreet}, J.~D. 2009, \araa, 47, 333

\bibitem[{{Donati} {et~al.}(1997){Donati}, {Semel}, {Carter}, {Rees}, \&
  {Collier Cameron}}]{Donaetal97}
{Donati}, J.-F., {Semel}, M., {Carter}, B.~D., {Rees}, D.~E., \& {Collier
  Cameron}, A. 1997, \mnras, 291, 658

\bibitem[{{Farihi} {et~al.}(2011){Farihi}, {Dufour}, {Napiwotzki}, \&
  {Koester}}]{Farietal11}
{Farihi}, J., {Dufour}, P., {Napiwotzki}, R., \& {Koester}, D. 2011, \mnras,
  413, 2559

\bibitem[{{Ferrario} {et~al.}(2015){Ferrario}, {de Martino}, \&
  {G{\"a}nsicke}}]{Ferretal15}
{Ferrario}, L., {de Martino}, D., \& {G{\"a}nsicke}, B.~T. 2015, \ssr, 191, 111

\bibitem[{{Gianninas} {et~al.}(2011){Gianninas}, {Bergeron}, \&
  {Ruiz}}]{Gianetal11}
{Gianninas}, A., {Bergeron}, P., \& {Ruiz}, M.~T. 2011, \apj, 743, 138

\bibitem[{{Grunhut} {et~al.}(2010){Grunhut}, {Wade}, {Hanes}, \&
  {Alecian}}]{Grunetal10}
{Grunhut}, J.~H., {Wade}, G.~A., {Hanes}, D.~A., \& {Alecian}, E. 2010, \mnras,
  408, 2290

\bibitem[{{Hussain}(2012)}]{Huss12}
{Hussain}, G.~A.~J. 2012, Astronomische Nachrichten, 333, 4

\bibitem[{{Kawka} \& {Vennes}(2012)}]{KawkVenn12}
{Kawka}, A. \& {Vennes}, S. 2012, \mnras, 425, 1394

\bibitem[{{Kepler} {et~al.}(2013){Kepler}, {Pelisoli}, {Jordan}, {Kleinman},
  {Koester}, {K{\"u}lebi}, {Pe{\c c}anha}, {Castanheira}, {Nitta}, {Costa},
  {Winget}, {Kanaan}, \& {Fraga}}]{Kepletal13}
{Kepler}, S.~O., {Pelisoli}, I., {Jordan}, S., {et~al.} 2013, \mnras, 429, 2934

\bibitem[{{Koester} {et~al.}(1998){Koester}, {Dreizler}, {Weidemann}, \&
  {Allard}}]{Koesetal98}
{Koester}, D., {Dreizler}, S., {Weidemann}, V., \& {Allard}, N.~F. 1998, \aap,
  338, 612

\bibitem[{{Koester} {et~al.}(2009){Koester}, {Voss}, {Napiwotzki},
  {Christlieb}, {Homeier}, {Lisker}, {Reimers}, \& {Heber}}]{Koesetal09}
{Koester}, D., {Voss}, B., {Napiwotzki}, R., {et~al.} 2009, \aap, 505, 441

\bibitem[{{Landstreet} {et~al.}(2014){Landstreet}, {Bagnulo}, \&
  {Fossati}}]{Landetal14}
{Landstreet}, J.~D., {Bagnulo}, S., \& {Fossati}, L. 2014, \aap, 572, A113

\bibitem[{{Landstreet} {et~al.}(2012){Landstreet}, {Bagnulo}, {Valyavin},
  {Fossati}, {Jordan}, {Monin}, \& {Wade}}]{Landetal12}
{Landstreet}, J.~D., {Bagnulo}, S., {Valyavin}, G.~G., {et~al.} 2012, \aap,
  545, A30

\bibitem[{{Landstreet} {et~al.}(2015){Landstreet}, {Bagnulo}, {Valyavin},
  {Gadelshin}, {Martin}, {Galazutdinov}, \& {Semenko}}]{Landetal15}
{Landstreet}, J.~D., {Bagnulo}, S., {Valyavin}, G.~G., {et~al.} 2015, \aap,
  580, A120

\bibitem[{{Landstreet} {et~al.}(1989){Landstreet}, {Barker}, {Bohlender}, \&
  {Jewison}}]{Landetal89}
{Landstreet}, J.~D., {Barker}, P.~K., {Bohlender}, D.~A., \& {Jewison}, M.~S.
  1989, \apj, 344, 876

\bibitem[{{Leone}(2007)}]{Leone07}
{Leone}, F. 2007, \mnras, 382, 1690

\bibitem[{{Liebert} \& {Stockman}(1980)}]{LiebStoc80}
{Liebert}, J. \& {Stockman}, H.~S. 1980, \pasp, 92, 657

\bibitem[{{Mathys}(1989)}]{Math89}
{Mathys}, G. 1989, \fcp, 13, 143

\bibitem[{{Petit} {et~al.}(2013){Petit}, {Owocki}, {Wade}, {Cohen},
  {Sundqvist}, {Gagn{\'e}}, {Ma{\'{\i}}z Apell{\'a}niz}, {Oksala}, {Bohlender},
  {Rivinius}, {Henrichs}, {Alecian}, {Townsend}, {ud-Doula}, \& {MiMeS
  Collaboration}}]{Petietal13}
{Petit}, V., {Owocki}, S.~P., {Wade}, G.~A., {et~al.} 2013, \mnras, 429, 398

\bibitem[{{Putney}(1997)}]{Putn97}
{Putney}, A. 1997, \apjs, 112, 527

\bibitem[{{Schmidt} \& {Smith}(1995)}]{SchmSmit95}
{Schmidt}, G.~D. \& {Smith}, P.~S. 1995, \apj, 448, 305

\bibitem[{{Tout} {et~al.}(2008){Tout}, {Wickramasinghe}, {Liebert}, {Ferrario},
  \& {Pringle}}]{Tout08}
{Tout}, C.~A., {Wickramasinghe}, D.~T., {Liebert}, J., {Ferrario}, L., \&
  {Pringle}, J.~E. 2008, \mnras, 387, 897

\bibitem[{{Valyavin} {et~al.}(2005){Valyavin}, {Bagnulo}, {Monin}, {Fabrika},
  {Lee}, {Galazutdinov}, {Wade}, \& {Burlakova}}]{Valyetal05}
{Valyavin}, G., {Bagnulo}, S., {Monin}, D., {et~al.} 2005, \aap, 439, 1099

\bibitem[{{Valyavin} \& {Fabrika}(1999)}]{ValyFabr99}
{Valyavin}, G. \& {Fabrika}, S. 1999, in Astronomical Society of the Pacific
  Conference Series, Vol. 169, 11th European Workshop on White Dwarfs, ed.
  S.-E. {Solheim} \& E.~G. {Meistas}, 206

\bibitem[{{Valyavin} {et~al.}(2008){Valyavin}, {Wade}, {Bagnulo}, {Szeifert},
  {Landstreet}, {Han}, \& {Burenkov}}]{Valyetal08}
{Valyavin}, G., {Wade}, G.~A., {Bagnulo}, S., {et~al.} 2008, \apj, 683, 466

\bibitem[{{Wade} \& {MiMeS Collaboration}(2015)}]{Wadeetal15}
{Wade}, G.~A. \& {MiMeS Collaboration}. 2015, in Astronomical Society of the
  Pacific Conference Series, Vol. 494, Physics and Evolution of Magnetic and
  Related Stars, ed. Y.~Y. {Balega}, I.~I. {Romanyuk}, \& D.~O. {Kudryavtsev},
  30

\bibitem[{{Wickramasinghe} {et~al.}(2014){Wickramasinghe}, {Tout}, \&
  {Ferrario}}]{Wicketal14}
{Wickramasinghe}, D.~T., {Tout}, C.~A., \& {Ferrario}, L. 2014, \mnras, 437,
  675

\bibitem[{{Zuckerman} {et~al.}(2011){Zuckerman}, {Koester}, {Dufour}, {Melis},
  {Klein}, \& {Jura}}]{Zucketal11}
{Zuckerman}, B., {Koester}, D., {Dufour}, P., {et~al.} 2011, \apj, 739, 101

\end{thebibliography}



\end{document}